\documentclass[sigconf,authorversion]{acmart}

\AtBeginDocument{%
  \providecommand\BibTeX{{%
    \normalfont B\kern-0.5em{\scshape i\kern-0.25em b}\kern-0.8em\TeX}}}
\usepackage{balance}
\usepackage{breakurl}
\usepackage[]{multibib}
\newcites{BI}{Appendix: References for Broader Impact Statements}
\newcites{reg}{References}

\setcopyright{none}
\acmYear{2021}
\acmConference[AIES '21]{Proceedings of the 2021 AAAI/ACM Conference on AI, Ethics, and Society}{May 19--21, 2021}{Virtual Event, USA}
\acmBooktitle{Proceedings of the 2021 AAAI/ACM Conference on AI, Ethics, and Society (AIES '21), May 19--21, 2021, Virtual Event, USA}
\acmDOI{10.1145/3461702.3462608}

\begin{document}
\fancyhead{}
\title{Unpacking the Expressed Consequences of AI Research in Broader Impact Statements}

\author{Priyanka Nanayakkara}
\email{priyankan@u.northwestern.edu}
\affiliation{%
  \institution{Northwestern University}
  \city{Evanston}
  \state{Illinois}
  \country{USA}
}
\author{Jessica Hullman}
\email{jhullman@northwestern.edu}
\affiliation{%
  \institution{Northwestern University}
  \city{Evanston}
  \state{Illinois}
  \country{USA}
}
\author{Nicholas Diakopoulos}
\email{nad@northwestern.edu}
\affiliation{%
  \institution{Northwestern University}
  \city{Evanston}
  \state{Illinois}
  \country{USA}
}

\newcounter{daggerfootnote}
\newcommand*{\daggerfootnote}[1]{%
    \setcounter{daggerfootnote}{\value{footnote}}%
    \renewcommand*{\thefootnote}{$\dagger$}%
    \footnote[2]{#1}%
    \setcounter{footnote}{\value{daggerfootnote}}%
    \renewcommand*{\thefootnote}{\arabic{footnote}}%
    }

\begin{abstract}
The computer science research community and the broader public have become increasingly aware of negative consequences of algorithmic systems. In response, the top-tier Neural Information Processing Systems (NeurIPS) conference for machine learning and artificial intelligence research required that authors include a statement of broader impact to reflect on potential positive and negative consequences of their work. We present the results of a qualitative thematic analysis of a sample of statements written for the 2020 conference. The themes we identify broadly fall into categories related to how consequences are expressed (e.g., valence, specificity, uncertainty), areas of impacts expressed (e.g., bias, the environment, labor, privacy), and researchers' recommendations for mitigating negative consequences in the future. In light of our results, we offer perspectives on how the broader impact statement can be implemented in future iterations to better align with potential goals.
\end{abstract}

\ccsdesc[500]{Computing methodologies~Artificial intelligence}
\ccsdesc[500]{Social and professional topics~Codes of ethics}

\keywords{AI ethics, broader impacts, anticipatory governance, thematic analysis}

\maketitle

\section{Introduction}
Scientists and the broader public have long grappled with the scientist’s role in considering the societal consequences of their work. According to philosopher of science Heather~\citet{douglas2009science}, scientific thinking since the 1960s has tended to embrace the notion of a value-free ideal, limiting the extent to which scientists engage with non-epistemic social, ethical, or political values in the scientific process. Yet, however well-established, this value-free ideal has failed to address the many ways in which such values invariably infiltrate the scientific enterprise, including how a scientist might make changes to their research agenda based on potential societal consequences. 

While the idea of values in design and technology is hardly new~(e.g., \cite{winner1980artifacts,friedman1996value}), the broader computer science community has recently begun to make more concerted attempts to challenge the value-free ideal. In particular, computer scientists have called for researchers to consider the downstream consequences of their work as part of the peer-review process~\cite{hecht2018s}, formally integrating the act of reflecting on both positive and negative societal consequences into the scientific enterprise. Suggestions like this one are timely, as the computer science research community---as well as the broader public---is becoming increasingly aware of the ways in which deployed technologies have disproportionate negative impacts on marginalized communities~\cite{benjamin2019race,noble2018algorithms,eubanks2018automating} and pose significant costs to the environment~\cite{strubell2019energy,schwartz2020green,bender2021dangers}), and is demanding scrutiny to account for negative consequences~\cite{diakopoulos2016}.  

In 2020, the Conference on Neural Information Processing Systems (NeurIPS), a top-tier conference for machine learning (ML) research, required that authors submit a broader impact statement as part of each paper submission. As per official guidance, the statement was meant to include both positive and negative potential societal consequences. NeurIPS’s broader impact requirement mirrored the call put forth by~\citet{hecht2018s}, and as a result of some ambiguity in its messaging has been termed by~\citet{hullman_blogpost} and others as an ``experiment,’’ ostensibly in facilitating more active and intentional engagement within computer science around societal consequences of research and technology.

One way to conceive of the act of writing the broader impact statement is as an ``ethical tool,’’ as defined by \citet{moula_sandin}; that is, ``a practical method and/or conceptual framework with the main purpose of helping the user(s) improve their ethical deliberations in order to reach an ethically informed judgement or decision.” Similarly, the broader impact statement is relevant to the ongoing conversation around Algorithmic Impact Assessments (AIAs)~\cite{moss2020governing}, which refer to methods of increasing accountability around algorithmic systems. More broadly, such impact statements may contribute to frameworks of Responsible Research and Innovation (RRI)~\cite{schomberg2013, owen2012} which help govern the R\&D process in ways that are responsive to ethical and societal concerns.

While the intended outcomes of the broader impact statement, as envisioned by the NeurIPS conference organizers, are ambiguous, the \citet{hecht2018s} proposal suggests goals such as increased transparency of impacts for the community, and encouragement of reflection and research on ways to mitigate negative impacts. In this paper we examine the content of NeurIPS broader impact statements in light of these goals. Do broader impact statements capture a wide array of positive and negative consequences? Is there evidence that authors are considering ways to mitigate negative impacts? We present a qualitative thematic analysis of hundreds of NeurIPS 2020 broader impact statements, characterizing the impacts---both positive and negative---and recommendations for mitigating negative consequences that researchers discussed in their statements. Our underlying goal is to gain insight into what and how researchers elaborated in their broader impact statements, with an eye toward how our results may inform similar future initiatives.
\section{Background}
The broader impact statements we study operate as a governance tool within the peer review process. Thus, we first briefly situate our work within ideas of responsible research before connecting to literature on values in science and technology.

\subsection{Responsible Research}
RRI can be understood as a process in which stakeholders ``become mutually responsible to each other and anticipate research and innovation outcomes"~\cite{schomberg2013}. This in turn relates to models of anticipatory governance in which emerging technologies are steered in order to be adapted to societal needs and ethical considerations~\cite{guston2014understanding}. In general, there are a range of approaches to RRI and anticipatory governance including scenario-based methods which can help systematically explore potential outcomes, and participatory methods which encourage the inclusion of societal stakeholders and feedback into the research process ~\cite{Brey:tl}. In this work we focus on impact assessments as a mode of anticipatory governance that allows for evaluation and engagement with the ethical implications of a given technology. For example, \citet{wright2011framework} introduces a framework for ethical impact assessments which includes five guiding ethical principles/dimensions (e.g., Respect for Autonomy, Nonmaleficence, etc., which relate to earlier work by \citet{beauchamp2001principles}) as well as an accompanying set of ethical tools and procedures to consider a technology as it relates to the guiding dimensions.

AIAs, as previously mentioned, are impact assessments specifically within an algorithmic context. \citet{moss2020governing} outline several remaining questions about the details of how AIAs work in practice. It is possible to imagine broader impact statements functioning as an ethical tool within a larger AIA framework. However, it is also important to note that the broader impact statement alone, as defined by NeurIPS, does not explicitly require wider engagement with the public or communities that are likely to experience harms that result from a certain technology's deployment or use. \citet{metcalf2021algorithmic} point out that this could result in abstract discussions of impacts that do not reflect impacts that are likely to be realized, suggesting that the statement may work better as one potential approach among other ethical tools used to surface relevant ethical issues around a given technology.

\subsection{Values in Technology}
In addition to philosophical inquiry into the role of values in science~\cite{douglas2009science,brown:2020}, there is a substantial body of work that lays the foundation for discussing the values embedded in technology. This includes work such as \citet{winner1980artifacts}'s re-envisioning of common artifacts as political ones, work by \citet{friedman1996value} and others on value-sensitive design, an approach to design that is closely intertwined with a consideration of values that may be built in to a technology, and \citet{barbrook1996californian}'s writing on the combination of values of the New Left and the New Right into a particular kind of ``Californian Ideology'' manifested in Silicon Valley's technological innovations.

There is also a growing body of work that specifically examines values within the computer science community, including how  those values are ultimately reflected in new technologies. \citet{rogaway2015moral} describes values that are implicit specifically within cryptographic work, and how these values may have shifted since cryptography's origins. \citet{hanna2020against} question the value of scalability in computer science, while \citet{birhane2020underlying} provide a study of prominent values found in the ML literature. In terms of ethical deliberation around technology, \citet{shen2020value} propose ``Value Cards,'' a toolkit intended to facilitate deliberation around technology and societal values. Within the specific domain of artificial intelligence (AI), recent attention has been given to value alignment challenges and whether values should be embedded or learned from data~\cite{russell:2019}.
\section{The NeurIPS 2020 Broader Impact Statement}\label{background}
In this section, we offer context on the NeurIPS 2020 broader impact statement requirement as background to our study. Authors submitting papers to NeurIPS in 2020 were required to also submit a broader impact statement, which would not count toward the official page limit. The official requirement from the call for papers\footnote{https://neurips.cc/Conferences/2020/CallForPapers} is sparse, including the following instructions: \textit{``In order to provide a balanced perspective, authors are required to include a statement of the potential broader impact of their work, including its ethical aspects and future societal consequences. Authors should take care to discuss both positive and negative outcomes."}

While the Frequently Asked Questions (FAQ) page\footnote{https://neurips.cc/Conferences/2020/PaperInformation/NeurIPS-FAQ} explains that a submission could not ``be rejected solely on the basis of the Broader Impact Statement," the call for papers states that ``[s]ubmissions will also be considered on ethical grounds'' and that ''a submission may be rejected for ethical considerations'' The conflict between these statements invites confusion around the role of the broader impact statement in determining acceptance status~\cite{hullman_blogpost}. Beyond the broader impact statements, the conference included an additional ethical review process, through which peer reviewers could flag papers for ethical concerns which would lead to thorough ethical reviews of these papers by designated ethics reviewers~\cite{lin2020what}. While official communication regarding ethics-related rejections is limited, it is plausible to think that broader impact statements could have played a role in the paper flagging process and subsequent ethics review. 

Official guidance for writing the broader impact statement was limited, but included a note on writing statements on theoretical work that the authors believe do not have foreseeable societal impact. In particular, the FAQ page states that authors can write in their statement that ``This work does not present any foreseeable societal consequence.'' Additionally, the conference webpage provides links to the call for broader impact statements to be incorporated into the peer-review process by ~\citet{hecht2018s}, two unofficial guides for statement writing~\cite{hecht2020suggestions,ashurst2020guide}, and a NeurIPS paper published in 2019~\cite{gillick2019breaking} which included a concluding statement very similar to a broader impact statement. While \citet{hecht2020suggestions}'s guide provides high-level suggestions, such as to begin writing the broader impact statement early, \citet{ashurst2020guide}'s guide provides authors with guiding questions, an ``Impact Stack'' to help facilitate systematic reflection on the layers of impact of a given technology, and multiple example statements. More broadly, \citet{prunkl2021institutionalizing} discuss benefits, risks, and challenges of a NeurIPS-like broader impact statement, and describe how the NeurIPS broader impact statement fits within a larger context of ethics practices (e.g., IRBs).

The program chairs of 2020's conference provide preliminary details~\cite{lin2020what} on the outcomes of the new broader impact requirement. In particular, they write that ``about 9\% of the submission[s] did not have such a section, and most submissions had a section with about 100 words.'' It is important to note that even though some submissions omitted the broader impact section, all camera-ready accepted papers were required to include it. \citet{abuhamad2020like} conducted a small preliminary survey ($N=50$) of researchers, the majority of whom submitted to NeurIPS 2020, and found that the authors surveyed expressed ``nonchalance'' toward the broader impact statement and saw it as a ``burden,'' although the majority also spent less than two hours on composing the statement itself. This research also found that there was confusion around how statements would be assessed, and that authors would have liked example statements as guidance. Additionally, \citet{boyarskaya2020overcoming} conducted an initial analysis ($N=35$) of NeurIPS 2020 broader impact statements that were available in paper pre-prints before publication of the conference pre-proceedings. They identified potential limitations of the broader impact statements in order to inform recommendations for future methods of anticipating harms.
\section{Methodology}
In this section we describe the data and analytic approach to our qualitative analysis of broader impact statements. 
\subsection{Data}
We first created a dataset including the 1,899 papers in the pre-proceedings of NeurIPS 2020. For each of these papers, we recorded the title, author name(s), and link to the full-length PDF by parsing \url{https://papers.nips.cc/paper/2020}. We then randomized the order of papers and extracted broader impact statements from the PDFs as needed for the analysis. We used regular expressions to extract all text after the heading for ``Broader Impact'' and before a heading for either ``Acknowledgements'' or ``References''; this method typically extracted the text for the broader impact statement of a given paper. In cases where the automated extraction method failed (e.g., when our parser did not account for a specific variation in heading titles), we manually recorded the text of the corresponding broader impact statement. These extracted texts formed the corpus\footnote{\url{https://docs.google.com/spreadsheets/d/1S3YAsFeuc0Aex_2geA7nD-qAm7FR8Lcu4eoMmEjUPVk/edit?usp=sharing}} that we used for subsequent analysis. 

\subsection{Analysis}
We qualitatively analyzed broader impact statements using a thematic analysis approach~\cite{Braun:2008dw,Gibbs2007}. We began by sequentially open coding excerpts from the randomly ranked set of collected impact statements. We allowed themes to emerge from the data and followed the method of constant comparison to continuously assess the fit and organization of excerpts into stable conceptual groupings that were triangulated across statements. To code, the first author read statements while pulling out excerpts that fit into existing thematic clusters of statements, or excerpts that could potentially be the basis for new themes. Throughout the process, the first author considered whether a particular theme could be broken into multiple themes, or whether themes could be re-organized to make more conceptual sense. The first author also wrote memos to articulate and clarify intermediate understandings of themes as well as relationships between themes. 

We also practiced analyst triangulation~\cite{Patton:2014} in which all co-authors periodically met to discuss the conceptual organization of themes and whether excerpts falling under particular themes fit under said themes, updating the overall organization accordingly. After coding 300 statements we felt themes were largely saturated and stable with little new information by each next statement~\cite{glaser2017discovery}. At this point, we stopped coding additional statements due to diminishing returns. We then revisited all 300 statements and tabulated the prevalence of selected themes. 
\section{Results}
We find that authors primarily use the broader impact statement to discuss consequences of their work and to at times offer suggestions for how to mitigate potential negative consequences. Because these topics naturally arise in the statements, we present our findings as they relate to these two broad categories: Impacts and Author Recommendations.

\subsection{Impacts}
In this section, we focus on features of broader impact statements that relate to potential consequences of research. Thus, we begin by describing \textit{how} authors present potential consequences (Expression of Impacts). Then we describe \textit{what} main categories emerged in terms of the types of impacts around which clusters formed (Types of Impacts), and the relationship authors describe between theoretical work and foreseeable societal consequences (Relationship to Theoretical Work). Finally, we present \textit{who} authors say will be impacted by their work (Who is Impacted?) and \textit{when} authors suggest society might see the consequences of their work (Timeframe of Impacts). We provide counts for selected themes beginning in Types of Impacts.

\subsubsection{Expression of Impacts.}
Authors express impacts along multiple axes. Specifically, we identify dimensions describing the \textit{valence}, \textit{orientation}, \textit{specificity}, \textit{uncertainty}, and \textit{composition} of stated consequences.

\textbf{\textit{Valence.}} 
We find that statements range in the extent to which they describe positive or negative consequences. We later describe themes that arise around valence with respect to composition, including patterns in how authors structure statements with regard to positive/negative consequences (see \textit{Composition}).

\textbf{\textit{Orientation.}}
We find that the areas of impact authors describe vary in terms of their orientation toward more society-facing outcomes (e.g., increased surveillance) versus more technically-oriented outcomes (e.g., increased robustness of a system). Technically-oriented consequences often have potential impacts on society (e.g., increased robustness of a system used in the context of self-driving cars could reduce the number of errors made, which keeps people safe), however authors do not always explicitly note such connections.

\textbf{\textit{Specificity.}} Authors express consequences with a range of specificity. At the more specific end of the spectrum, authors say their work can be used in particular ways in specific contexts such as education (e.g., “this work will be integrated as part of the tools that we intend to teach in Ph.D. courses”~\citeBI{nguyen2020distributionally}\daggerfootnote{Reference is part of our NeurIPS 2020 broader impact statement dataset and can be found in the Appendix.}). Authors also point to specific areas of research that their work will impact, specific ways in which their work will contribute to a certain line of research, or a specific type of person who will be impacted. In at least a few cases, authors provide detailed, domain-specific case studies describing how their work may impact society (e.g., ~\citetBI{coston2020counterfactual}$^{\dagger}$ describe their work in the context of parole decisions), or give examples of how the completed work may have already impacted society (e.g., the environmental impacts of having trained neural network models~\citeBI{jain2020interpretable}$^{\dagger}$).

In terms of very general consequences, authors sometimes name broad impact areas (e.g., ``This work primarily has applications in the ﬁelds of psychophysics and crowdsourcing, and more generally, in learning from human responses.'' ~\citeBI{perrot2020near}$^{\dagger}$) but don't always explain how the work will impact those areas. Finally, some authors express the idea that societal outcomes are dependent on the use case, or are application-specific, so societal consequences are difficult to discuss.

\textbf{\textit{Uncertainty.}}
Authors contend with the limitations of their methods, and the uncertainty inherent in models, pointing these out in statements in the vein of negative consequences. These consequences are more technically-oriented, with gestures toward associated societal impacts. Some authors point out that if assumptions are not met or adequately considered, the system could lead to wrong or harmful conclusions (e.g.,~\citetBI{moreno2020robust, squires2020active}$^{\dagger}$). Additionally, authors acknowledge uncertainty around models by pointing out model assumptions. \citetBI{hao2020profile}$^{\dagger}$ note that “[a] potential downside is that the theoretical guarantees of the associated algorithms rely on the assumption correctness.” 

Authors are also careful to point out the lack of guarantees provided by their algorithms, saying that “it is also important for the buyer to beware; \ldots there may remain cases of interest when our approximations fail without warning”~\citeBI{ghosh2020approximate}$^{\dagger}$ or that “we cannot guarantee that every constraint will hold”~\citeBI{narasimhan2020approximate}$^{\dagger}$ as well as other general limitations, such as that “density-ratio based methods are not a panacea; it is entirely possible for the technique to introduce new biases when correcting for existing ones”~\citeBI{rhodes2020telescoping}$^{\dagger}$.

Finally, authors describe how there is a possibility of errors downstream in future training or deployment, and that the real-world behavior of a given system is uncertain. For example, \citetBI{kumar2020one}$^{\dagger}$ write that “it can be difficult for a practitioner to predict the behavior of an algorithm when it is deployed” and ~\citetBI{hoang2020revisiting}$^{\dagger}$ write that “[w]hile applications of [their] work to real data could result in ethical considerations, this is an indirect (and unpredictable) side-effect of [their] work.” Some authors describe specific ways in which there could be downstream errors. ~\citetBI{mate2020collapsing}$^{\dagger}$ write that “if the worker makes an error when entering data \ldots then the algorithm could make the wrong recommendation” and ~\citetBI{bhatia2020preference}$^{\dagger}$ write that if a designer chooses an inappropriate target, it “could lead to incorrect inferences and unexpected behavior in the real world.”

\textbf{\textit{Composition.}}
The composition dimension reflects the overall conceptual structure and mixture of material in a statement in terms of valence, orientation, specificity, and uncertainty.

Some statements contain a mixture of valences, including both positive and negative consequences (though not necessarily balanced equally). We also find statements that lean heavily toward positive consequences (that is, they mostly describe positive consequences but mention negative consequences briefly). On one extreme, we see that some authors \textit{only} include positive consequences, omitting a discussion of negative consequences entirely. While some simply don’t make any mention of negative impacts, others explicitly state that there are no negative consequences (e.g., “Our work does not have any negative societal impacts” ~\citeBI{wu2020firefly}$^{\dagger}$; “no one is put at disadvantage from this research” ~\citeBI{dinh2020consistent}$^{\dagger}$).

In addition, we find a theme around the statement of negative consequences where some authors write that a particular technology could be misused by a malicious actor, leading to undesirable societal outcomes. In this sense, authors engage with uncertainty around how a technology could be used in the future. \citetBI{scialom2020coldgans}$^{\dagger}$ provide an example of potential misuse:

\begin{quote}
\ldots malicious actors can use the same technology to build tools detrimental to society, e.g. for creation and propagation of misleading (fake) news \ldots, impersonation, and deceit.
\end{quote}

In addition, we find some statements structured such that they predominantly reflect positive consequences, but include a brief caveat alluding to potential negative consequences; the caveat often appears near the end of the statement. \citetBI{zhou2020promoting}$^{\dagger}$ describe the positive impacts of their work, then conclude with the following caveat: ``However, any reinforcement learning method runs the risk of being applied to military activities. Our work is no exception.''

We also find that some statements do not contain a discussion of societal consequences at all. The omission of consequences is typically stated simply: “This work does not present any foreseeable societal consequence,” mirroring the instruction provided in the call for papers. In these cases, no reason is provided for the omission of consequences. In other cases, authors provide reasons around why there may be limited consequences of their work. For example, some authors write that because their work is based on existing work, their contributions do not significantly change what is technologically possible. Along these lines, \citetBI{biggs20203d}$^{\dagger}$ write about potential negative impacts of their work, but limit the extent to which the work introduces new types of impacts:

\begin{quote}
Furthermore, our method is an improvement of existing capabilities, but does not introduce a radical new capability in machine learning. Thus our contribution is unlikely to facilitate misuse of technology which is already available to anyone.
\end{quote}

Finally, we find that authors' statements are sometimes oriented solely toward technical consequences, only describing their work in the context of how it contributes to the research space. In these cases, the stated technical contributions are usually framed as positive consequences.

\subsubsection{Types of Impacts.}
Our analysis finds several categories of impacts that authors tend to discuss in broader impact statements. Attributes of statements that fall into these categories vary in terms of the aforementioned dimensions of expression (e.g., \textit{Composition}, \textit{Valence}, etc.). In particular, we point out that these categories range in terms of their orientation (see Expression of Impacts: \textit{Orientation}) toward societal versus technical consequences. For each type of impact presented below, we note the category's predominant orientation as it appeared in statements.

\textbf{\textit{Privacy.}} Impacts around privacy ($19.3\%, N=58$) are largely oriented toward societal outcomes including protection of personal data and increased surveillance. In terms of positive impacts, authors describe how their work facilitates local computation (therefore reducing the extent to which personal data is shared) (e.g., “we help facilitate on-device deep learning, which could replace traditional cloud computation and foster the protection of privacy”~\citeBI{chao2020directional}$^{\dagger}$), contributes to methods that rely on decentralization (e.g., “One beneﬁt of decentralized algorithms is that it does not need the central node to collect all users’ information and every node only communicates with its trusted neighbors”~\citeBI{liu2020decentralized}$^{\dagger}$), and contributes to other privacy-focused research areas such as differential privacy (e.g., \citetBI{chen2020understanding}$^{\dagger}$) which aim to further protect an individual's information.

On the other hand, authors note how their work could be detrimental to maintaining privacy. They write that their work could facilitate the identification of individuals (e.g.,“potential use cases may target users based on their activity patterns”~\citeBI{rambhatla2020provable}$^{\dagger}$), contribute to surveillance methods (e.g., “some common concerns on [Deep Neural Networks (DNNs)] such as privacy breaching and heavy surveillance can be worsened by DNN devices that are more available economically by using our proposed techniques”~\citeBI{lee2020flexor}$^{\dagger}$), or even ``may raise issues in neuroethics, especially regarding mental privacy''~\citeBI{jain2020interpretable}$^{\dagger}$.

\textbf{\textit{Labor.}}
We find that authors write about societally-oriented impacts of their research in terms of influences on employment/work ($6.3\%, N=19$) and impacts to productivity. For example, they acknowledge that automation of manually-done processes could displace jobs. \citetBI{mate2020collapsing}$^{\dagger}$ write the following on the jobs of community health workers (CHWs):

\begin{quote}
\ldots we also present results that highlight our algorithm’s ability to plan among
thousands of processes at a time, far more than for which a human could independently plan. Just making this capability available could encourage the automation of applicable interventions via automated calls or texts, potentially displacing CHW jobs \ldots
\end{quote}

\citetBI{chao2020directional}$^{\dagger}$ write more generally that ``unemployment may increase due to the increased automation enabled by the deep learning.” On the other hand, some authors state that while their work contributes to automation, it is intended to increase productivity (e.g., ``These techniques are not meant to replace highly skilled human workers, but to help improve their productivity at work”~\citeBI{he2020geo}$^{\dagger}$). Some authors explicitly describe the tension between automation reducing jobs versus increasing productivity:

\begin{quote}
Like many AI technologies, when used in automation, our technology can have a positive impact (increased productivity) and a negative impact (decreased demand) on labor markets.~\citeBI{van2020mdp}$^{\dagger}$
\end{quote}

Additionally, authors write about how their work could improve or facilitate work processes such as collaboration, for instance “by helping bridge the gap between the ML and the scientific computing communities through allowing them to share tools and more easily interoperate”~\citeBI{moses2020instead}$^{\dagger}$ or by helping to bring multiple disciplines together (e.g., by introducing concepts from one field (or subfield) into another~\citeBI{yang2020bayesian, zhou2020learningmanifold}$^{\dagger}$). Authors also describe how their work could be useful for decision-making tasks. In particular, they write about decision-making in the contexts of medicine and policy, and how their work could contribute to fairer decisions. Along these lines, ~\citetBI{hu2020fair}$^{\dagger}$ write the following:

\begin{quote}
By adopting our method, decision makers can build multiple decision models simultaneously just from one historical dataset and ensure that all decision models will be fair after the deployment \ldots
\end{quote}

\textbf{\textit{Environment.}}
We find that authors describe societally-oriented impacts to the environment ($10\%, N=30$). For instance, authors describe how decreases in computational costs for a given task that their work accomplishes could ``[save] lots of energy consumption"~\citeBI{cheng2020hierarchical}$^{\dagger}$ and ``has the potential to reduce the carbon footprint of building AI models"~\citeBI{daulbaev2020interpolation}$^{\dagger}$. In terms of other positive consequences, they write about how their work could be used to contribute to climate change models:

\begin{quote}
By employing Gaussian processes for data assimilation and building them into larger frameworks, this could facilitate more accurate climate models compared to current methods \ldots~\citeBI{borovitskiy2020matern}$^{\dagger}$
\end{quote}

On the other hand, authors also note that the amount of energy necessary to implement their methods could be harmful to the environment:

\begin{quote}
\ldots the carbon footprint of the proposed approach remains low compared to many deep approaches used nowadays in machine learning applications \ldots~\citeBI{jaquier2020high}$^{\dagger}$
\end{quote}

\textbf{\textit{Media.}}
Some authors seem to be aware of the impact their work could have on the media ($6\%,N=18$). They especially focus on impacts around generating or spreading misinformation, through synthetic media such as images or video (i.e., deepfakes) or other means. In this way, impacts around media are quite societally-oriented. For example, ~\citetBI{phuoc2020blockgan}$^{\dagger}$ draw attention to how their work could be used in service of fake news: ``Similar to existing image editing software,
this enables the creation of image manipulations that could be used for ill-intended misinformation (\textit{fake news}).''

Not all authors discussing misinformation, however, describe their work as aiding the spread of misinformation. At least one set of authors describes how their work could mitigate the harmful effects of misinformation, as it ``may be used to detect harmful articles which may contain fake news, violent content, and fraudulent materials"~\citeBI{wang2020bidirectional}$^{\dagger}$.

\textbf{\textit{Bias.}} When it comes to impacts around bias (including fairness, [$24\%, N=72$]), authors write both about impacts in ways that are both societally and technically-oriented. We note that definitions of bias vary across these two orientations, and even within an orientation, sometimes take on slightly different meanings. We first describe impacts around bias that is expressed as societally-oriented and maps to historical forms of discrimination. We then describe how authors sometimes refer to bias in more technically-oriented ways---in these cases, they do not connect bias of a system to societal discrimination.

Bias that can manifest in discriminatory algorithmic decisions is frequently mentioned in broader impact statements. In particular, authors engage with the idea of biased datasets, which may refer to training data populations in which some groups are underrepresented (see \cite{suresh2019framework}), though few authors define what they mean by dataset bias in their statements. Some write that biased datasets are not an issue in their work; for instance, \citetBI{zhang2020counterfactual}$^{\dagger}$ write that they ``validate [their] method on large-scale public vision-language datasets and do not leverage biases in the data.'' Others describe how biases in datasets (stemming from society at large) may play a role in their work or the impacts of their work:

\begin{quote}
\ldots because the majority of [genome-wide association studies] are performed on individuals of
European ancestries, [polygenic scores] are more accurate for individuals from those ancestry groups, potentially exacerbating health disparities between individuals of different ancestries \ldots~\citeBI{spence2020flexible}$^{\dagger}$
\end{quote}

Additionally, authors write about discriminatory bias in the context of models. For instance, \citetBI{liu2020generalized}$^{\dagger}$ write that the signal recovery model of interest (which takes as input a separate model) could inherit potential biases of the input model. \citetBI{zhang2020self}$^{\dagger}$ describe the impacts of bias potentially introduced during training:

\begin{quote}
\ldots our ﬁnding advocates for
the use of priors for the regularization of neural networks. Despite the potentially better generalization performance of trained models, depending on the choice of priors used for training, unwanted bias can be inevitably introduced into the deep learning system, potentially causing issues of fairness \ldots
\end{quote}

On the other hand, authors talk about how their work could be used to mitigate these biases in algorithmic systems or datasets. For example, \citetBI{chen2020self}$^{\dagger}$ describe how their work helps address dataset bias:

\begin{quote}
\ldots our theoretical work guides efforts to mitigate dataset bias. We demonstrate that curating a diverse pool of unlabeled data from the true population can help combating existing bias in labeled datasets. We give conditions for when bias will be mitigated and when it will be reinforced or ampliﬁed by popular algorithms~\ldots
\end{quote}

At times, authors write about impacts around bias in a technical sense. For example:

\begin{quote}
First, we note that finding clusters in networks is critical to \textit{reducing bias} on measuring interventions with network effects \ldots~Having more flexible and better ways of doing this clustering will improve our ability to assess treatments on networks. \citeBI{liu2020strongly}$^{\dagger}$
\end{quote}

It is, however, not always clear whether authors are referring to bias in a societal or technical sense, or whether technical forms of bias are related to societal inequities. To illustrate this point, we point to \citetBI{xie2020noise2same}$^{\dagger}$'s description of how failure in their system could lead to biases in downstream systems:

\begin{quote}
On the negative aspect, as many imaging-based research tasks and computer vision applications may be built upon the denoising algorithms, the failure of [the paper's framework for denoising] could potentially lead to biases or
failures in these tasks and applications. 
\end{quote}

The authors do not specify the form of bias they have in mind, which underscores an aspect of the dimension of specificity in the impact statements. In cases such as this, terminology could be understood in different ways and has not been precisely articulated and disambiguated.

\textbf{\textit{Efficiency.}}
Efficiency ($30\%, N=90$) is a predominantly technically-oriented attribute that appears to often be prioritized in these research contexts, though justifications for it vary if given at all. However, authors do sometimes map the impact of increased/decreased efficiency to societal outcomes. For instance, authors write about how the increased efficiency afforded by their work (e.g., in terms of decreased training time) could save either time, money, or other resources. As an example, ~\citetBI{zhou2020learningimplicit}$^{\dagger}$ write about how their work could be applied to achieve ``efficient package delivery with swarms of drones to reduce delivery costs.” In addition, authors write about how increased efficiency demonstrated by their work could be beneficial for the environment, in terms of reduced computational costs, an idea that we see in the \textit{Environment} theme as well. Some authors connect increased efficiency to the democratization of technology (see the \textit{Democratization} theme below). For example, \citetBI{larsson2020strong}$^{\dagger}$ write that “the improved efficiency resulting from the predictor screening rules will make the [Sorted L-One Penalized Estimation (SLOPE)] family of models \ldots accessible to a broader audience, enabling researchers and other parties to fit SLOPE models with improved efficiency” and go on to note that their work makes it possible to analyze ``data sets that were otherwise too large to be analyzed without access to dedicated high-performance computing clusters.” 

While some authors imply that efficiency is a positive outcome, they may not explicitly draw a connection between increased efficiency and a particular societal outcome of efficiency. In these cases, efficiency as an impact functions in very much of a technically-oriented capacity. For example, \citetBI{brennan2020greedy}$^{\dagger}$ write about improving the effectiveness of a certain method, then write that they ``hope to make [the method] more tractable, efficient, and broadly applicable"; because they don't make an explicit connection between efficiency itself and a particular outcome, it is difficult to make conclusions about the reasoning behind efficiency in particular. Finally, there are cases where authors invoke the role of efficiency in enabling other impacts such as those related to privacy or increased surveillance (e.g., “there are various unexpected yet feasible negative collateral consequences of increased efficiency, e.g., in terms of privacy”~\citeBI{alvarez2020geometric}$^{\dagger}$).

\textbf{\textit{Generalizability.}} Authors write about the generalizability of their methods ($15.3\%, N=46$) as positive consequences of their work, orienting generalizability technically, but sometimes connecting the technical contribution to a potential societal consequence. While it seems generally accepted that generalizability leads to positive outcomes (or is positive in and of itself), \citetBI{hu2020one}$^{\dagger}$ highlight the tension between generalizability and potential application to societally undesirable use cases. In particular, they write that ``[their] method has the potential to generalize the existing AI algorithms to more applications, while it also raises serious concern of privacy” going on to acknowledge the potential negative consequences that could arise from applying their work to the online behaviors of web users.

Authors also write about scalability as an important outcome or limitation of their work. For example, \citetBI{yang2020bayesian}$^{\dagger}$ note that while they improve scalability, their improvements are not quite enough, as it is still not possible to ``scale to environments with thousands of agents and hundreds of types where the state-action space has thousands of dimensions.''

\textbf{\textit{Democratization.}} A theme emerges where researchers see their work as democratizing ($5.7\%, N=17$) by making some technology available to more people or organizations (e.g., in cases where previously using the technology required significant resources). Thus, we find that authors primarily characterize the increase in democratization of technology as a positive consequence of their work. In particular, they write about how their work enables more people in both research and non-research contexts to make use of various technologies. In this way, impacts around accessibility are both technically-oriented (e.g., improving researchers' accessibility to a particular technology enables further research in a given area) and societally-oriented (e.g., increasing accessibility of a particular technology enables wider use and participation by the wider public). More specifically, authors write about providing researchers with the ability to “harness volunteer computing and train models on the scale currently available only to large corporations”~\citeBI{ryabinin2020towards}$^{\dagger}$, while also discussing how their work can “help all the amateurs to use machine learning without any hassle”~\citeBI{zhang2020differentiable}$^{\dagger}$. As more of an exception, \citetBI{hong2020low}$^{\dagger}$ write directly about both the potential positive and negative consequences of increasing accessibility of technology:

\begin{quote}
Techniques like [Spatially Stochastic Networks (SSNs)] that make content creation more accessible can help bridge this gap in representation [of developers] \ldots~As any technology that promises to give easier access, it has the potential for misuse. One could imagine cases where generative models are used to create things that may be harmful to society, and this can lower the technical entry barrier to misusers of this technology. For instance, SSNs could be applied toward harmful DeepFakes.
\end{quote}

\textbf{\textit{Robustness and Reliability.}} Authors describe impacts around robustness (e.g., against adversarial attacks) and reliability ($21.3\%, N=64$). In particular, they cite robustness as a positive attribute of their work, or lack of robustness as a limitation. While impacts in this area tend to be more technically-oriented, authors do draw connections to more society-facing impacts that stem from robustness. They note the importance of robustness in technologies that are used in contexts where safety is of concern, such as in ``self-driving cars’’~\citeBI{diakonikolas2020complexity}$^{\dagger}$. Along the same lines of minimizing system errors, authors also write about their contributions with respect to increased reliability, including the ways in which their work can help people contend with uncertainty, noting that ``for real-world classiﬁcation tasks like disease diagnosis, in addition to accurate predictions, we need reliable estimates of the level of conﬁdence of the predictions made”~\citeBI{zhang2020self}$^{\dagger}$.

\textbf{\textit{Accuracy.}} We find that the concept of accuracy $(12.7\%,N=38)$ arises in statements. Authors write about improvements to accuracy as positive consequences of their work, mainly as technically-oriented consequences with allusions to societal consequences. For instance, they write that their work ``could have significant broader impact by allowing users to more accurately solve practical problems”~\citeBI{hoang2020revisiting}$^{\dagger}$ or more simply stating that their ``method \ldots can lead to even higher accuracies” which ``is of high practical relevance”~\citeBI{masegosa2020second}$^{\dagger}$.

\textbf{\textit{Interpretability.}} Authors describe impacts of their work that relate to interpretability ($10.3\%, N=31$). For instance, authors write about how their contributions to increased interpretability of algorithms or models can contribute to understanding biases or “promoting fairness in artificial intelligence”~\citeBI{yue2020interventional}$^{\dagger}$:

\begin{quote}
Moreover our algorithms have the additional benefit that they lead to more readily interpretable hypotheses \ldots~This could potentially help practitioners better understand and diagnose complex machine learning systems they are designing, and troubleshoot ways that the algorithm might be amplifying biases in the data.~\citeBI{chen2020classification}$^{\dagger}$
\end{quote}

Authors also write about how interpretability relates to the potential consequence of engendering trust in systems, which is framed as a positive consequence:

\begin{quote}
In a positive prospect, we believe that our model contributes to further the development of less opaque machine learning models \ldots debugging and interpreting the model’s behaviour and can help to establish trust toward the model when employed in larger application pipelines.~\citeBI{ehrhardt2020relate}$^{\dagger}$ 
\end{quote}

In other cases, authors cite lack of interpretability as a negative consequence (or limitation) of their work. For example,~\citetBI{zhou2020learningimplicit}$^{\dagger}$ say that “[their] method \ldots faces the `black box problem’ where behaviors of the individual agents may not be rational or interpretable from the human perspective.” They go on to talk about potential fairness issues, suggesting a connection between lack of interpretability and potential for societally negative outcomes. 

\subsubsection{Relationship to Theoretical Work.}
We find that authors can differ in how they frame the relationship between theoretical work and societal implications. A theme emerges where authors indicate, either explicitly or implicitly, that due to the theoretical nature of their work, there are \textit{no} forseeable, negative, or ethical consequences ($9\%, N=27$). In particular, some narrow down the claim to be that there are no direct \textit{negative} impacts of the work. However, this does not necessarily mean that they omit a discussion of other impacts of the work (positive or technical). Sometimes authors of theoretical work write that there are no foreseeable consequences altogether due to the nature of their contributions:

\begin{quote}
The current paper presents theoretical work without any foreseeable societal consequence. Therefore, the authors believe that the broader impact discussion is not applicable.~\citeBI{emek2020stateful}$^{\dagger}$
\end{quote}

Note that these statements are slightly different from those that only state there are no foreseeable consequences (see Expression of Impacts: \textit{Composition}), as they make a connection between theoretical work and lack of foreseeable consequences. We see another theme where authors write about the difficulty of foreseeing societal consequences of their work (e.g., ``We believe that it is difﬁcult to clearly foresee societal consequence of the present, purely theoretical, work''~\citeBI{luneau2020information}$^{\dagger}$).

Other authors mention the theoretical nature of their work, and write that there may be indirect consequences of their work or describe potential societal consequences ($10\%, N=30$). For example:

\begin{quote}
Our results, while being theoretical in nature, have potential impacts in providing instructions for designing better security tools to ensure that people’s online activities do not create unintended leakage of private information.~\citeBI{tang2020optimal}$^{\dagger}$
\end{quote}

\subsubsection{Who is Impacted?}
In more than half of the broader impacts statements we analyzed, authors reference not only impacts, but also who might be impacted ($64.3\%, N=193$). These references vary in specificity. In a broad sense, authors often cited a domain (e.g., healthcare) which motivates their work or could be positively impacted by their work (e.g.,~\citetBI{pinheiro2020unsupervised}$^{\dagger}$ write about the need for labeled data in medical imagery applications, and \citetBI{yang2020factorizable}$^{\dagger}$ write about how their work could help with ``discovering the reasons for the quick spread of the epidemic disease in some areas”). At times, authors are more specific, mentioning particular groups who could be positively impacted, such as “people with language deficits like aphasia”~\citeBI{jain2020interpretable}$^{\dagger}$ or people who could benefit from “[r]etinal prostheses”~\citeBI{mahuas2020new}$^{\dagger}$. We also find a cluster of statements which mention impacts relating specifically to members of creative industries. For example, \citetBI{tancik2020fourier}$^{\dagger}$ write that “this progress may also inadvertently reduce employment opportunities for the human artists who currently produce [photorealistic] effects manually,” which fits into another theme of authors noting that those whose jobs may be automated could be negatively impacted (see Types of Impacts: \textit{Labor}). ~\citetBI{xie2020noise2same}$^{\dagger}$ write about their work on deep image denoising saying that “[i]ndividuals and corporations related to photography may benefit from [their] work," offering a potential positive consequence for a creative industry.

Other times, authors focus on impacts to members of various slices of the general public, such as those who use social media. For example, \citetBI{zhu2020fewer}$^{\dagger}$ say that because “[their] work can be used in social media companies or any other occasions where user data can be accessed, people who are worried about their privacy being analyzed or targeted may be put at disadvantage.” Authors also note how their work could positively or negatively impact people from historically disadvantaged groups, for example “by preventing them from receiving biased decisions”~\citeBI{hu2020fair}$^{\dagger}$ or by being used “to monitor and negatively impact individuals in protected groups.”

Lastly, authors turn their attention to researchers and practitioners. They write about how their work could benefit researchers both within computer science (e.g., ``[o]ur study will \ldots benefit the machine learning community”~\citeBI{liu2020improved}$^{\dagger}$) and outside of computer science (e.g., atmospheric sciences~\citeBI{berner2020numerically}$^{\dagger}$). We assume that when authors write about how their work could open up future research, they are implying that other researchers will be impacted. Authors also write that their work can help practitioners in their processes, such as by assisting them in ``formally [reasoning] about the trade-offs across accuracy, robustness, and communication efficiency”~\cite{chen2020distributed}$^{\dagger}$. 

\subsubsection{Timeframe of Impacts.}
When authors write about the timeframe of impacts, that is, when society may see the consequences of the work at hand, they use broad terms that generally refer to a timeframe in terms of immediate versus long-term impact, and how their work might accelerate the realization of some outcome ($10.3\%, N=31$). For instance, they refer to work as having “no immediate practical applications”~\citeBI{meulemans2020theoretical}$^{\dagger}$ or work that “can in the long run cause positive or negative societal impact”~\citeBI{van2020mdp}$^{\dagger}$. The level of detail in terms of timeframe typically does not go beyond immediate versus non-immediate results. 

\subsection{Author Recommendations}
While authors tend to focus more attention on impacts, we also find that they sometimes make recommendations for how to mitigate potential negative consequences. We describe outcomes that authors suggest working toward, including their proposed steps for achieving these outcomes and their ascriptions about responsibility for taking action.

\subsubsection{Outcomes.}
We describe five categories of desirable future outcomes that authors write about for the field and society to consider. In several cases, these outcomes align with positive impacts of their completed work, as described in Impacts: \textit{Types of Impacts}.

\textbf{\textit{Safe and Effective Use of AI.}}
Authors make suggestions for how to ensure the safe or otherwise more effective use of their work, or AI in general $(20.7\%, N=62)$. In particular, they suggest that ``the practitioner verify how realistic these modeling assumptions are for the application at hand’’~\citeBI{alvarez2020geometric}$^{\dagger}$ and note the importance of ``[understanding] the principle and limitation of an algorithm to prevent failure”~\citeBI{ok2020graph}$^{\dagger}$. Authors also suggest more work around model robustness and write about the need for interpretability of models (e.g., with regards to algorithms used in safety-critical environments~\citeBI{xie2020deep}$^{\dagger}$). Additionally, authors suggest human-in-the-loop methods (e.g., to check outputs in a personalized medicine context~\citeBI{huang2020dense}$^{\dagger}$), the need to develop methods of detecting misuses of technology (e.g., in the case of deepfakes~\citeBI{shan2020meta}$^{\dagger}$), and the potential need for regulation or policy around certain technologies (e.g., again in the case of deepfakes~\citeBI{hong2020low}$^{\dagger}$).

\textbf{\textit{Ensure ``Fair" Outcomes.}}
With regards to recommendations around issues of fairness ($6.3\%, N=19$), authors sometimes make general mentions of the importance of paying future attention to “fairness and nondiscrimination”~\citeBI{liao2020provably}$^{\dagger}$ or the area of algorithmic fairness as it relates to their work~\citeBI{nguyen2020distributionally}$^{\dagger}$. Other times, authors make more specific suggestions, such as recommending to diversify tasks or datasets, saying, for instance, that “it is important to wisely choose the data on which the system is trained \ldots~If data is biased our method is not guaranteed to provide a correct estimation; this could harm the final users and should be carefully taken into account”~\citeBI{patacchiola2020bayesian}$^{\dagger}$.

\textbf{\textit{Protect Privacy.}}
Authors make recommendations to protect privacy ($5.3\%, N=16$). For example, authors make recommendations to ``deploy [their] model locally"~\citeBI{shan2020meta}$^{\dagger}$ and nod toward the importance of privacy-focused research areas such as differential privacy (e.g.,~\citetBI{nguyen2020distributionally}$^{\dagger}$) moving forward.

Additionally, authors write more generally about the need for “privacy protections \ldots throughout data collection, training, and deployment”~\citeBI{adhikari2020learning}$^{\dagger}$, and broadly refer to legislative or regulatory means as ways of protecting privacy going forward. \citetBI{foster2020adapting} write that ``[they] welcome regulatory efforts to produce a legal framework that steers the usage of machine learning \ldots in a direction which \ldots respects \ldots the privacy rights of users.''

\textbf{\textit{Reduce Environmental Impact.}}
Similar to how authors bring up increased efficiency of their work as potentially benefiting the environment, they make recommendations for increased efficiency as a means of reducing environmental impact in the future ($1\%, N=3$).

\subsubsection{Who is Responsible?}
We find that researchers sometimes make recommendations but do not assign responsibility ($7.3\%, N=22$), leaving it unclear who is ultimately responsible to take follow-up action. However, when authors do make mention of who might be responsible (either explicitly or implicitly) ($24\%, N=72$), they sometimes refer to various stakeholders, including people involved closer to deployment (e.g., practitioners, system designers, field experts, companies). They also refer to researchers (sometimes authors say that they are personally responsible for some action, like being involved in policy discussions around a certain technology; others leave their suggestions as a call to action for the broader research community; others describe recommendations for future computer science work implying that the research community is responsible) and policymakers.

\section{Discussion and Conclusion}
We find salient themes that emerge in broader impact statements from NeurIPS 2020; these findings indicate that in aggregate, authors discussed a fairly wide range of consequences including mentioning ideas for mitigating negative consequences. These descriptive findings contribute a set of dimensions and themes which could inform guidelines to shape what and how a research community might frame a broader impact statement requirement so as to reduce ambiguity of intention that might have contributed to the wide range of themes we observed, such as by guiding expectations around authors' treatment of orientation, specificity, uncertainty, range of impacts, who is impacted, timeframe of impacts, and so on. 

In closing, we elaborate a discussion based on how the statement appears to function in light of our results. We consider our findings through the lens of three potential goals: encouraging reflexivity, initiating changes to future research, and minimizing negligence and recklessness. Each offers opportunities for research communities such as NeurIPS to further steer or even evaluate broader impact statements.

\textbf{\textit{Encouraging Reflexivity.}} One potential goal of the statement might have been simply to encourage researchers to reflect on the societal consequences of their work, or engage in moral inquiry, which \citet{bietti2020ethics} argues is intrinsically valuable. More formally, reflexivity as a political rationality found in the literature on social responsibility is ``focused on making scientists aware of their own values and motivations, as well as making them reflect on the possible outcomes of their scientific inquiries''~\cite{glerup2014mapping}. 

Our analysis finds evidence that the exercise surfaced values of the computer science research community, through the relatively frequent mentions of themes like efficiency, robustness, generalizability, or privacy, but it is less clear whether researchers devoted much effort to deliberating on their own personal values. In this sense, our findings contribute to the recent investigation into values of ML research by \citet{birhane2020underlying}, who identify common values present in influential ML publications such as accuracy, efficiency, and generalization. However, while they find that ``overt consideration of societal benefits or harms is extremely rare,'' we find that in aggregate, broader impact statements do address societal consequences, suggesting that the statement requirement helps address an otherwise overlooked area. We did also observe some statements that reference technical contributions but do not necessarily map these contributions to potential \textit{societal} impacts. If the goal is to encourage reflection on societal impacts, it could be useful for the official requirements/guidance around the statement to further encourage taking impacts an extra step further to consider how society could be impacted as a result of technical contributions (see \citet{ashurst2020guide}'s ``Impact Stack'') and to also encourage taking into account context when considering potential harms~\cite{boyarskaya2020overcoming}.

\textbf{\textit{Initiate Changes to Future Research.}} Second, a potential goal of the broader impact statement might have been to help initiate changes in research agendas; the idea being that upon considering societal consequences of their work, researchers may have determined the potential negative consequences to far outweigh potential positive ones, prompting a shift in future research. Because significant time has not passed since authors wrote the statement, it is difficult to know whether it played any role in shifting research directions. It could be useful to observe whether content of statements shifts over time, and if discussions that arise in one year's set of statements manifest themselves in the work done in the following year(s). It could also be useful to conduct interviews with researchers to understand how, if at all, writing the statement informed consequent research questions. 

\citet{hecht2020suggestions} suggests that writing a broader impact statement early on in the research process leaves room to adapt a project accordingly, which is also in line with ideas from anticipatory governance \cite{Brey:tl}. Our analysis finds that researchers most likely completed their broader impact statements toward the end of the research process, namely because statements describe impacts of the work in more of a retroactive manner. In part, this could have been because the announcement of the broader impact requirement came at a time when work that was ultimately submitted was already later-stage. If the goal is to allow earlier-stage and continuous reflection, the requirements for the statement could benefit from clarification or perhaps changes in implementation to ensure researchers draft statements (or engage in ethical deliberation) early on in their research processes.

\textbf{\textit{Minimize Negligence and Recklessness.}} Third, we explore the possibility that the broader impact statement may function as a way to minimize research negligence and recklessness. According to \citet{douglas2009science}, ``[r]ecklessness is proceeding in the face of unreasonable risk; negligence is the failure to foresee and mitigate such risk.'' The broader impact statement, therefore, can be seen as ensuring that authors engage with potential consequences, therefore reducing the chance of negligence. By making suggestions for mitigating negative consequences, researchers also minimize the risk of recklessness, in some way contributing to the mitigation of the risks of their own work by offering potential ways forward. While our findings already show a significant breadth of impacts, it could nonetheless be helpful to provide authors with particularly relevant ethical issues that relate to their general area of work, such as by referring them to established ethics tools (e.g.,~\cite{wright2011framework}), so that they may do a more complete job of assessing risks.

Finally, we note that several of the above perspectives imply that a practice like requiring broader impacts statements can be successful without drastic changes aimed at increasing the skills and knowledge of computer scientists around ethics and anticipation of societal outcomes of technology. For example, if the statement is intended to minimize negligence and recklessness, this would seem to imply that the community as a whole is largely capable of identifying the potentially harmful consequences of their work. Some have questioned, however, whether computer scientists are prepared to make useful observations about potential ethical problems with their work, and whether the results will be closer to ``speculative fiction''~\cite{adarblog}. Additionally, participatory ethics methods \cite{Brey:tl} suggest that community stakeholders likely to be affected should be involved in research to ensure that researchers are aware of possible problems they might not foresee. In the long run, it would seem that any expected instrumental values of broader impacts will depend on a clearly defined stance on the extent to which the community is intended to seek new skills, or collaborations, in reflecting on ethical implications. 

\begin{acks}
Thank you to the Navigating the Broader Impacts of AI Research Workshop for providing a space for discussion about the broader impact statement, and to Jack Bandy and the reviewers for their helpful comments.
\end{acks}

\bibliographystyle{ACM-Reference-Format}
\balance
\bibliography{ref.bib}

\appendix
\bibliographystyleBI{ACM-Reference-Format}
\balance
\bibliographyBI{ref.bib}

\end{document}